\def\year{2022}\relax
\documentclass[letterpaper]{article} % DO NOT CHANGE THIS
\usepackage{aaai22}  % DO NOT CHANGE THIS
\usepackage{times}  % DO NOT CHANGE THIS
\usepackage{helvet}  % DO NOT CHANGE THIS
\usepackage{courier}  % DO NOT CHANGE THIS
\usepackage[hyphens]{url}  % DO NOT CHANGE THIS
\usepackage{graphicx} % DO NOT CHANGE THIS
\usepackage{natbib}  % DO NOT CHANGE THIS AND DO NOT ADD ANY OPTIONS TO IT
\usepackage{caption} % DO NOT CHANGE THIS AND DO NOT ADD ANY OPTIONS TO IT
\DeclareCaptionStyle{ruled}{labelfont=normalfont,labelsep=colon,strut=off} % DO NOT CHANGE THIS
\usepackage{algorithm}
\usepackage{algorithmic}

\usepackage{amsmath}
\usepackage{amsfonts}

\usepackage[table,xcdraw]{xcolor}

\usepackage{newfloat}
\usepackage{listings}
\floatstyle{ruled}
\newfloat{listing}{tb}{lst}{}
\floatname{listing}{Listing}
\title{CTMSTOU driven markets: simulated environment for regime-awareness in trading policies}
\author{
    Selim Amrouni,
    Aymeric Moulin,
    Tucker Balch
}

\affiliations{
    J.P. Morgan AI Research\\
    383 Madison Ave\\
    New York, New York 10017\\
   \{selim.amrouni; aymeric.ca.moulin; tucker.balch\}@jpmorgan.com
}

\usepackage{bibentry}
\begin{document}

\maketitle

\begin{abstract}

Market regimes is a popular topic in quantitative finance even though there is little consensus on the details of how they should be defined. They arise as a feature both in financial market prediction problems and financial market task performing problems.

In this work we use discrete event time multi-agent market simulation to freely experiment in a reproducible and understandable environment where regimes can be explicitly switched and enforced.

We introduce a novel stochastic process to model the fundamental value perceived by market participants: Continuous-Time Markov Switching Trending Ornstein-Uhlenbeck (CTMSTOU), which facilitates the study of trading policies in regime switching markets.

We define the notion of \emph{regime-awareness} for a trading agent as well and illustrate its importance through the study of different order placement strategies in the context of order execution problems.

\end{abstract}

\section{Introduction}
\label{sec:intro}
Financial Markets have been a topic of interest for decades. Researchers have studied prediction of future market states as well as carrying out tasks like execution or hedging in the current market state. Authors often refer to the the notion of market regime. Markets are believed to show remarkably different properties in different states. These regimes are sometimes described as lasting years to months in macro economic contexts or rather minutes to hours in micro structure contexts. There is currently no widely accepted definition of market regime. There are multiple ways to describe them. Typically, current levels of  volatility, price direction, bid-ask spread, volume traded can be used to describe a regime. In this work we just seek to show the potential of regime awareness and stick to a simple upward trending vs downward trending regime distinction for simplicity.

We do not focus on predicting future states of the market (e.g. next day price prediction) but rather focus on carrying out tasks in a market. Here we develop on the task of algorithmic execution. It consists in buying or selling a given quantity that is typically too big to be executed in one single order. The quantity needs to be split into several smaller orders executed on some time horizon. The decision making is typically thought of in two parts: scheduling and order placement model. Scheduling: decide on the target quantity execution profile as a function of time. Order placement model: decide whether to place market order or limit order and the limit price that should be used for limit orders. The general intuition on the problem is that executing the entire order on a short time window is safer in terms of market risk: the stock price has less time to move which means that the average price of execution is closer to the initial price. However doing this typically requires using more market orders and increases the execution cost by creating a short term liquidity shortage and potential adversarial behaviors in the market. On the other hand executing on a longer schedule typically decreases execution cost but increases the risk associated with market price movement. The choice on this trade-off on order placement model is made by considering risk appetite as well as market regime. In this work we illustrate the edge brought by regime awareness. We use the well established benchmark Time Weighted Average Price (TWAP) \cite{Berkowitz1988TheTC} for the quantity schedule and experiment with the order placement model.

To ensure freedom of experimentation, reproducibility and a better understanding we work with simulated markets. We build upon the multi-agent discrete event time market simulator ABIDES-Markets \cite{byrd2020abides}. ABIDES-Markets simulates a market where a single stock trades on a single exchange. The order book evolves with different types of market participant agents observing the state of the market and placing orders according to their strategies. The market configurations available aim at producing "stylized realism". The agent types are understandable (e.g. market maker, value agent, momentum agent, noise agent), their meta parameters and number of agents have been calibrated to decently match markets stylised facts \cite{vyetrenko2019real}.
The value agents play a key role in driving the overall price trends in the simulation. They place their orders based on their observation of a common fundamental value time series. If they are present in sufficient relative number with sufficient relative order size they will enforce the simulated mid price time series being close to the fundamental value time series. We experiment with the fundamental time series supplied to influence the market regime and create switches between price upward trending and downward trending regimes.

We propose a new stochastic process for the fundamental: Continuous-Time Markov Switching Trending Ornstein-Uhlenbeck (CTMSTOU). By design it switches between clearly identified growth regimes. Each regime is modeled by a trending Ornstein-Uhlenbeck process and the switches happen according to a continuous time Markov process.

\begin{figure}[t]
\centering
\includegraphics[width=0.9\columnwidth]{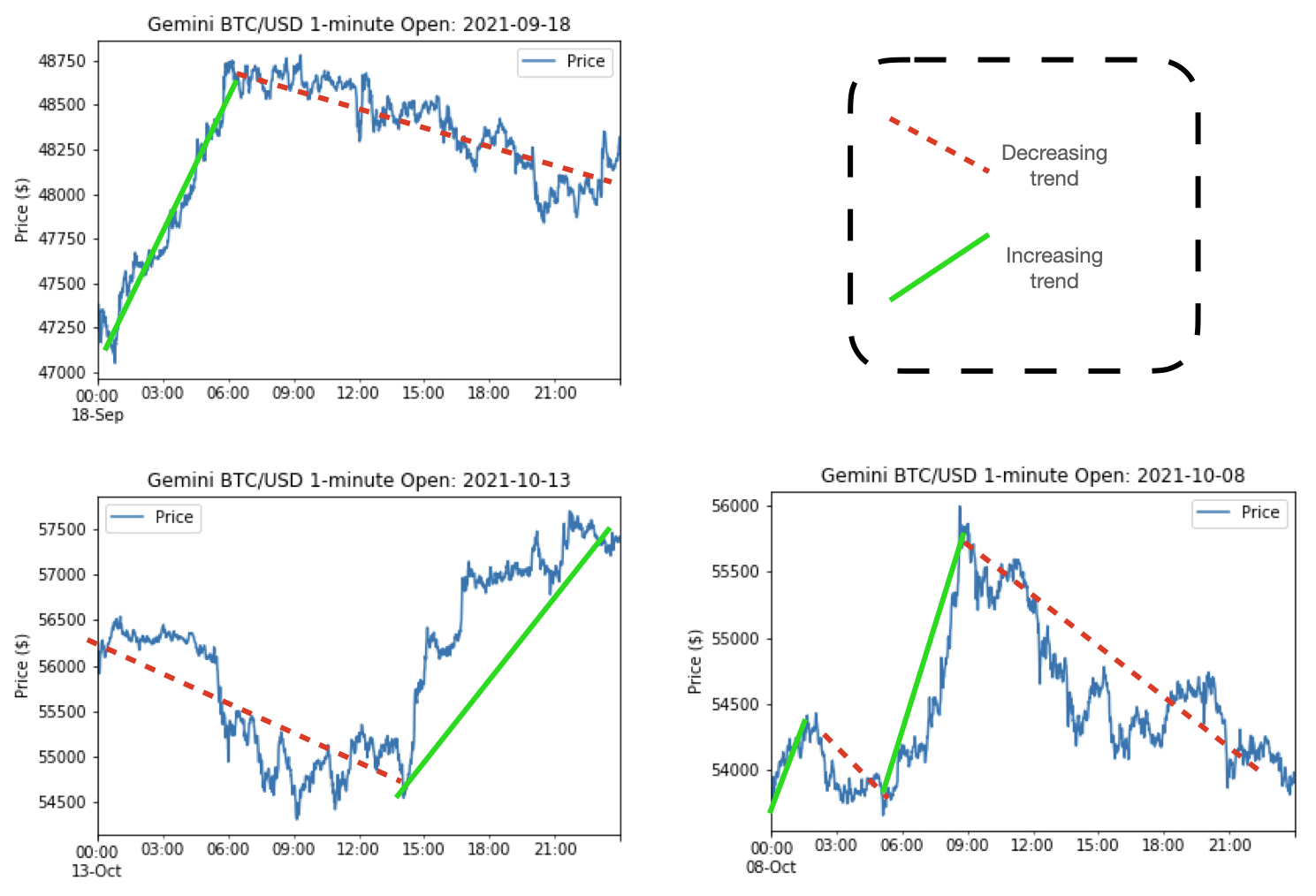} % Reduce the figure size so that it is slightly narrower than the column. Don't use precise values for figure width.This setup will avoid overfull boxes.
\caption{BTC/USD 1-minute bin Open price in the Gemini exchange. Examples of daily upward trends in  and downward trends in red.}
\label{fig:btc_usd_day_label}
\end{figure}

In practice we take the example of BTC/USD trading on the exchange Gemini. We use minute per minute aggregated data to calibrate our CTMSTOU process for the simulation fundamental. (Illustration on \ref{fig:btc_usd_day_label})

\section{Financial Markets Agent Based Simulation}
Agent based simulation is an important tool for research in quantitative trading. An important number of simulation platform have been developed such as \cite{8247866} or \cite{g12020046} and have reported promising results. As highlighted in \cite{amrouni2021abides}, ABIDES \cite{byrd2020abides} is a very flexible discrete time multi-agent event simulator with an already advanced extension to financial markets ABIDES-Markets. We choose to build upon ABIDES-Markets in this work.

    \subsection{Core simulation}
    At its core ABIDES is a discrete event time multi-agent simulator. Multiple agents take actions and communicate with each other using a messaging system. Each agent is awaken at the start, resulting in messages sent and delayed wake-up calls. These events are stored in a time priority queue and processed by the simulation kernel one by one. The simulation time jumps from an event time to another as the kernel process the event. Theoretically this functioning allows to simulate in a continuous time fashion. In practice the maximum time granularity supported is the nanosecond however two events with the same timestamp to the nanosecond will not be executed simultaneously, they will be executed one after the other based on arrival order in the queue.

    \subsection{Agents and Market Mechanics}
    ABIDES-markets implements a market with a single exchange and several market participants. The exchange is an agent in the simulation, it receives order messages and data request messages from market participants and sends them back acknowledgments and market data. The exchange is designed to obey NASDAQ continuous trading session rules \cite{nasdaq_market}. BTC/USD trading on Gemini exchange follows very similar rules. \cite{liang2019towards}.
    
    Market participant agents receive market data from the exchange, apply their trading policy to decide on which actions to take and send their instructions to the exchange. Most agents only see the world through their communication with the exchange (e.g. momentum agents, noise agents, \cite{DBLP:journals/corr/abs-1904-12066} gives a description of the agents). Only value agents have information outside of that. They have access to a stochastic process named fundamental. It aims to model the fundamental value of the asset perceived by investors.

    \subsection{Central Role of the Fundamental}
    As described above, the fundamental is the only exogenous factor in the market simulation. Without it, the overall price would be totally determined by the micro structure arising from agent policies. Value agents are the only agents that have access to the fundamental (direct access or to a noisy observation). They are inspired from zero-intelligence agents \cite{10.2307/2138676} and described in \cite{vyetrenko2019real}. Their strategy consists in placing limit orders with a price around the fundamental value and try to profit by capturing temporary deviations from the fundamental that revert. However it is the presence of these value agents itself that has the potential to enforce reversion to the fundamental value. A configuration with too few value agents (or with value agents that do not post enough shares relatively to the overall market posting rate) would exhibit very little reversion to the fundamental price and cause these agents to be unprofitable.
    
    Previous works in ABIDES-Markets typically either used historical data or mean reverting OU process. In the historical data approach the fundamental is a historical binned mid price time series of a stock in an effort to try to propose reactive simulation of trading that stock on that day. In the mean reverting OU process the parameters of the OU process parameters are fitted to mimic typical stock price behavior then for each simulation the fundamental time series is generating using the fitted stochastic process.
    
    Using either one of the these two approaches we could try to identify different regimes in the fundamental. Instead, to remove this layer of modeling and uncertainty we propose to introduce an adaptation of the fundamental with clearly distinct regimes with switches.

%\subsection{Method of calibration}
\section{Ornstein-Uhlenbeck Fundamental Adaptation}
    \subsection{Introduction}
    The original version of the Ornstein-Uhlenbeck (OU) process used for the fundamental in \cite{c:byrd2019explaining} is defined by the stochastic differential equation:
    \[
    dX_t = \theta\times (\mu - X_t) dt + \sigma dW_t 
    \]
    
    where~$\mu \in \mathbb{R}$ is the drift term, ~$\theta \in \mathbb{R}^{+}$, ~$\sigma \in \mathbb{R}^{+}$ and $W_t$ is a Wiener process.

    \subsection{Existing OU Processes Extensions}
        \subsubsection{Trending OU Process (TOU)}
        \cite{Thierfelder2015TheTO} proposes to add a linear with time trend component to the process that is defined: 
        \[ 
        d[X_t - \lambda t] = \theta\times (\mu - [X_t - \lambda t]) dt + \sigma dW_t 
        \] 
        where ~$\lambda \in \mathbb{R}$ is the trend term, ~$\mu \in \mathbb{R}$ is the drift term, ~$\mu \in \mathbb{R}^+$, ~$\sigma \in \mathbb{R}^{+}$ and $W_t$ is a Wiener process. Intuitively, once the trend term is removed from the process, the new process acts as a classical OU process. \cite{c:mejia2018} explains a methodology to calibrate this TOU process.
        
        \subsubsection{Functional Form OU Process (FOU)}
        \cite{dehling2017estimating} proposes to extend OU processes by replacing the constant drift term $\mu$ with a functional term $f(t)$:
        \[ 
        dx_t = \theta\times (f(t) - x_t) dt + \sigma dW_t 
        \] 
        where ~$\lambda \in \mathbb{R}$ is the trend term, ~$f(t) \in \mathbb{R}$ is the deterministic functional drift term, ~$\sigma \in \mathbb{R}^{+}$ and $W_t$ is a Wiener process. The resulting process is mean reverting around the time dependent center $f(t)$. This may be used to model seasonality or trends in the data.

    \subsection{Markov Switching Processes}
    \subsubsection{Idea for a time-step based simulation}
    TOU and FOU enable the modeling of financial price or fundamental belief time series with a trend. However, they do not enable modeling time series with structural breaks and brutal parameter changes. \cite{kuan2002lecture} proposes the use of Markov Switching models for ARCH(\cite{bollerslev1994arch}) and GARCH(\cite{bauwens2006multivariate} models. It enables multiple structures for parameters and a switching mechanism governed by a Markovian state variable. In such model, the quantity of interest $z_t$ evolves according to the following recurrence equations depending on the current state $s_t$: 
    
    \[
    z_t = \left\{
        \begin{array}{ll}
            \alpha_0 + \beta_0 z_{t-1} + \sigma_0 \epsilon_t & \mbox{if } \ s_t = 0 \\
            \alpha_1 + \beta_1 z_{t-1} + \sigma_1 \epsilon_t & \mbox{if } \ s_t = 1
        \end{array}
    \right.
    \] 
    where ~$\alpha_i \in \mathbb{R}$, ~$\beta_i \in \mathbb{R}$, ~$\sigma_i \in \mathbb{R}$ and ~$\epsilon_t \sim N(0,1)$
    
    \cite{10.2307/1912559} proposes to have $s_t \in [0, k],\ k \in N$ an unobservable state variable governed by a first order Markov Chain with transition matrix $\Pi$ such that $\Pi_{i,j} = Pr(S_t=i|S_{t-1}=j)$. Hence, the Markovian $s_t$
    variables result in random and frequent changes, the persistence of each regime depends on the transition probabilities and regime classification is probabilistic and determined by data.
    
    \subsubsection{Nuances in Discrete-Event simulation}
    To use the concept in discrete event time simulation we propose to adapt the Markov Switching process by using the Continuous-time Markov Chain (CTMC) framework \cite{anderson2011continuous}. The process switches state according to an exponential time random variable and then move to a different state as specified by the transition matrix. It is used to model a system with a set of states and where rates of changes from one state to another are known. The process has a kernel \cite{vadori2015semi} $Q$ of the form: $Q(i,j,t) = \Pi(i,j)(1-e^{-\lambda(i)t})$ where $\lambda(i) \in J$ are positive constants, $\Pi$ is a stochastic matrix (non negative entries and each row summing to one).
    
    \subsection{Proposed Adaptation: Continuous-Time Markov Switching Trending Ornstein-Uhlenbeck (CTMSTOU)}
    Aiming to develop an environment to study trend regime shifts in a discrete-event simulation, we propose to model the fundamental with what we introduce as a Continuous-Time Markov Switching Trending Ornstein-Uhlenbeck (CTMSTOU) process. This process is defined as: 
    \begin{equation}
    dX_t = \theta_{s_t} \times (M_{s_t} - X_t) dt + \sigma_{S_t} dW_t 
    \label{eq:final_process}
    \end{equation}
    where ~$s_t \in [\![0, k]\!],\ k \in \mathbb{N}^{*}$ is an unobservable state variable governed by a Continuous-Time Markov Chain with transition matrix $\Pi$ such that ~$\Pi_{i,j} = Pr(S_t=i|S_{t-1}=j)$, ~$M_{s_t} \in \mathbb{R}$ is the time dependent center term, ~$\sigma_n \in \mathbb{R}^{+}$, ~$n \in [\![0, k]\!]$ and $W_t$ is a Wiener process.  ~$M_{s_t}$ is defined as a function continuous and linear by piece such that $M_{s_0} = M_0, \ \frac{dM}{dt} = \mu_{s_t}$, ~$\mu_{s_t}$ is the slope of the function at time $t$, it can be seen as the trend of the market at time $t$. Finally, $M_t$ can be computed as: $M_t = M_0 + \int_{0}^{t}\mu_{s_t}dt$. In this particular case, because $\frac{dM}{dt}=constant$, we can compute $M_t= M_0 + \sum_{n=1}^{N_t} \mu_{T_{i-1}}\times (T_i - T_{i-1})$   
    
    \subsection{Generating Paths From CTMSTOU: Solving The Stochastic Process}
    A linear Stochastic Differential Equation (SDE) is written: 
    \begin{equation}
    dX_t = a(X_t,t) dt + b(X_t,t)  dW_t,\ X_0 = x_0
    \label{eq:general_sde}
    \end{equation}
     
    A solution is a stochastic process $X_t$, satisfying:
    \begin{equation}
    X_t = X_0 + \int_{0}^{t} a(X_t,t) dt + \int_{0}^{t} b(X_t,t)  dW_t
    \label{eq:general_solution_sde}
    \end{equation}
    
    For instance geometric brownian motion described by the SDE $dS_t = S_t\mu dt + S_t\sigma dW_t$  can be solved analytically: $S_t=S_0e^{\sigma W_t + (\mu - \frac{1}{2}\sigma^2)t}$. However, most SDEs do not have such solutions and we propose to use the Euler-Maruyama method \cite{higham2001algorithmic} for solving the equation \ref{eq:final_process}.
    
    \subsubsection{Euler-Maruyama Method}
    \label{sec:euler}
    For each step, we compute:
    \begin{equation}
    X_{n+1} = X_n + a(X_{t_n},t_n) \delta t + b(X_{t_n},t_n)  \Delta W_n
    \label{eq:euler_solution}
    \end{equation}
    where ~$\Delta W_n = W_{t_{n+1}} - W_{t_{n}}$, approximating ~$\int_{t_n}^{t_{n+1}} a(X_t,t) dt \approx a(X_{t_n},t_n) \delta t $, and ~$\int_{t_n}^{t_{n+1}} b(X_t,t)  dW_t \approx b(X_{t_n},t_n)  \Delta W_n $ because of $\delta t$ being small.
    % It is usual to use fixed time steps $\delta t$ as most of the simulation are discrete-time simulations. Because of the discrete-event property of the simulator ABIDES, we propose here a small modification of the Euler-Method:
    % \begin{algorithm}[tb]
    % \caption{Continuous-time Euler-Maruyama method}
    % \label{alg:algorithm_euler}
    % \textbf{Input}: $x_0$ the starting point, $\tau$ the parameter of the agent wakeup \\
    % \textbf{Parameter}: Optional list of parameters\\
    % \textbf{Output}: Your algorithm's output
    % \begin{algorithmic}[1] %[1] enables line numbers
    % \STATE Let $t=0$.
    % \WHILE{condition}
    % \STATE Do some action.
    % \IF {conditional}
    % \STATE Perform task A.
    % \ELSE
    % \STATE Perform task B.
    % \ENDIF
    % \ENDWHILE
    % \STATE \textbf{return} solution
    % \end{algorithmic}
    % \end{algorithm}

    \subsubsection{Visualization}
\begin{figure}[t]
    \begin{minipage}[c]{0.49\columnwidth}
    \includegraphics[width=\columnwidth]{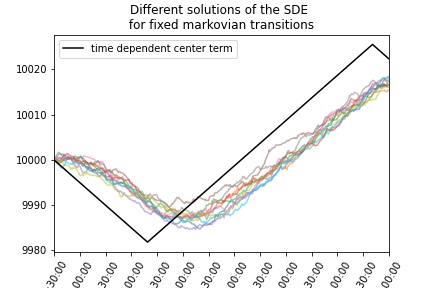}
    \end{minipage}
    \begin{minipage}[c]{0.49\columnwidth}
    \includegraphics[width=\columnwidth]{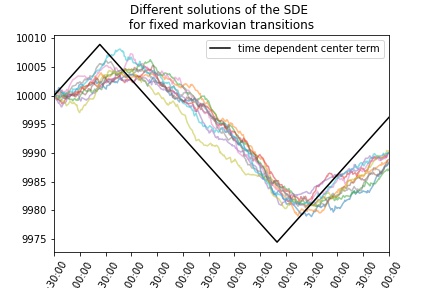}
    \end{minipage}
    \caption{CTMSTOU example: several sample paths for two different fixed center term series}
    \label{fig:example_OU}
    \end{figure}

    Here we define a CTMSTOU with the following parameters: ~$s_t \ \in \{1,2\}$, ~$x_0 = M_0 = 10000$, ~$\sigma_1 = \sigma_2 = 2$, ~$\theta_1 = \theta_2 = 1$, ~$\mu_1 = 10$ (upward market), ~$\mu_2 = -10$ (downward market) and ~$\Pi = \frac{1}{3600} \begin{pmatrix} \frac{1}{2} & \frac{1}{5}\\ \frac{1}{5} & \frac{1}{2} \end{pmatrix}$. Figure \ref{fig:example_OU} shows different sample paths from the solutions of the SDE computed using the Euler-Maruyama \ref{sec:euler} for a fixed Markovian transition between states. We represent different solution for the same fixed time dependent center term series $\{M_t\}$ only for illustration purpose. In practise in the simulation we only sample one path from the solution for a given center term series.
    
\begin{figure}[t]
    \begin{minipage}[c]{0.49\columnwidth}
    \includegraphics[width=\columnwidth]{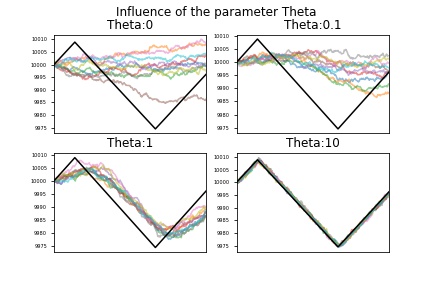}
    \end{minipage}
    \begin{minipage}[c]{0.49\columnwidth}
    \includegraphics[width=\columnwidth]{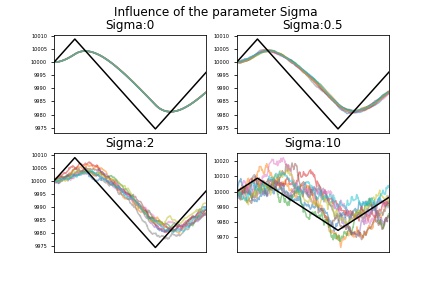}
    \end{minipage}
    \caption{Influence of parameter $\theta$ and $\sigma$}
    \label{fig:influence_thetha_sigma}
    \end{figure}

    \paragraph{Varying the parameters}
    We illustrate the impact of varying the parameters of the SDE:
    \begin{itemize}
        \item $\theta$: As illustrated in figure \ref{fig:influence_thetha_sigma}, all remaining parameters equal, the greater $\theta$ returns to the center term.
        \item $\sigma$: As illustrated in figure \ref{fig:influence_thetha_sigma}, and as we can expect, the greater $\sigma$, the more noisy is the process. 
    \end{itemize}
    
    % \begin{figure}[t]
    % \centering
    % \includegraphics[width=0.9\columnwidth]{figures/influence_theta.jpg} % Reduce the figure size so that it is slightly narrower than the column. Don't use precise values for figure width.This setup will avoid overfull boxes.
    % \caption{Influence of parameter $\theta$}.
    % \label{fig:influence_theta}
    % \end{figure}
    
    % \begin{figure}[t]
    % \centering
    % \includegraphics[width=0.9\columnwidth]{figures/influence_sigma.jpg} % Reduce the figure size so that it is slightly narrower than the column. Don't use precise values for figure width.This setup will avoid overfull boxes.
    % \caption{Influence of parameter $\sigma$}.
    % \label{fig:influence_sigma}
    % \end{figure}
    
    \subsection{Calibrating the Continuous-time Markov Transition Matrix}
    For the sake of simplicity we define two states: \emph{bullish} and \emph{bearish} \cite{clarke1998bullish} and as defined in \cite{bull_bear}:
    \begin{itemize}
        \item Bullish: upward trend in a particular asset (Bulls think the markets will go up)
        \item Bearish: downward trend in a particular asset (Bears think the market will go down)  
    \end{itemize}
    
    Hence, the transition matrix can be written $\Pi = \begin{pmatrix} \lambda_{11} & \lambda_{12}\\ \lambda_{21} & \lambda_{22} \end{pmatrix} $ where the $\lambda_{ij}$ are the rates of transition from $S_i$ to $S_j$ in event per second. So the bigger the $\lambda_{ij}$ the more likely the transition is. We additionally set $\lambda_{11} = \lambda_{22} = \lambda,\ \lambda_{12}=\lambda_{21}=\omega$ making the matrix symmetric and reducing the number of parameters to calibrate to two. We propose to calibrate $\Pi$ with the protocol described in algorithm \ref{alg:tr_mat}, this protocol is fairly simple and leaves room for further improvement.

        \subsubsection{Regime Switch Labelling on the Real Data}
        Algorithm \ref{alg:tr_mat} presents the general protocol to find the parameters $\lambda$ and $\omega$. Labelling and counting the regime switches in the real data is a critical step. We use data downloaded from \cite{crypto_data} which, among others, provides minute-aggregated OHLC data (Open High Low Close). We focused our analysis on 2021 data Gemini Exchange from 2021-01-01 to 2021-11-30 as illustrated in figure \ref{fig:btc_usd_day_label}. 
        
        % \begin{figure}[t]
        % \centering
        % \includegraphics[width=0.9\columnwidth]{figures/gemini_btc_usd_2021.png} % Reduce the figure size so that it is slightly narrower than the column. Don't use precise values for figure width.This setup will avoid overfull boxes.
        % \caption{Gemini BTC/USD in 2021}.
        % \label{fig:gemini_btc_usd_2021}
        % \end{figure}
        
         We use a methodology similar to \cite{srivastava2018evaluating} where a four regime models is computed. We compute the Moving Average (MA) of the bin open prices on the last 6-hours, compare it with the last 12-hours moving average and use it as classification feature. Volatility based bands \cite{cowie2011evaluation} based on the volatility of the last 6-hours of the Open price is used to avoid unnecessary regime switches (e.g., if the two MA are too close and produces false switching signals). The algorithm \ref{alg:tr_mat} describes the procedure used for regime labeling.
        
        \begin{figure}[t]
        \centering
        \includegraphics[width=0.8\columnwidth]{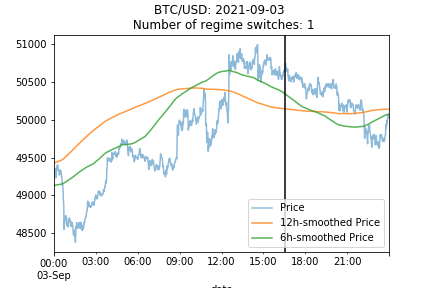} % Reduce the figure size so that it is slightly narrower than the column. Don't use precise values for figure width.This setup will avoid overfull boxes.
        \caption{Labelling example: Gemini BTC/USD on 2021-09-03. An observer may tell there are 2 regimes on this day, a roughly increasing trend during half of the day followed by a decreasing trend. The algorithm successfully picked up the 2 regimes (only 1 switch) even though the two MA curves cross 2 times. The volatility was important enough so that the algorithm did not pick the first switch.}
        \label{fig:btc_usd_day_label_2}
        \end{figure}
        
        \subsubsection{Find the parameters of the matrix}
        \label{section:find_parameters}
\begin{figure}[t]
        \centering
        \includegraphics[width=0.70\columnwidth]{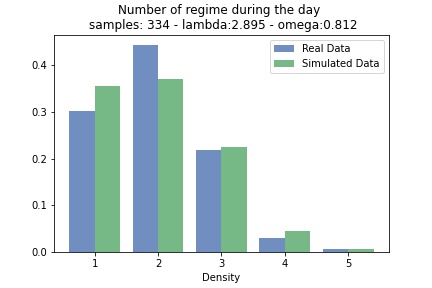} % Reduce the figure size so that it is slightly narrower than the column. Don't use precise values for figure width.This setup will avoid overfull boxes.
        \caption{Real and simulated distribution comparison  $\lambda = 2.90\ [event/day],\ \omega= 0.812\ [event/day]$}.
        \label{fig:gemini_btc_usd_2021}
        \end{figure}

        We propose to run a random search over the parameters $\lambda$ and $\omega$ using \cite{liaw2018tune} to minimize the distance between the distribution of regime switches count in the Gemini BTC/USD data and data simulated using the CTMSTOU. We use the Wasserstein distance \cite{ramdas2015wasserstein} to measure similarity between the simulated and real data. We find the values $\lambda = 3.356e^{-05}\ [event/sec] = 2.90\ [event/day],\ \omega  = 9.40e^{-06}\ [event/sec] = 0.812\ [event/day]$. The figure \ref{fig:gemini_btc_usd_2021} compares the distribution of the number of regimes throughout a day for real data and simulated data using these parameters.
        
    \subsection{Simulation output}
\begin{figure}[t]
    \centering
    \includegraphics[width=0.70\columnwidth]{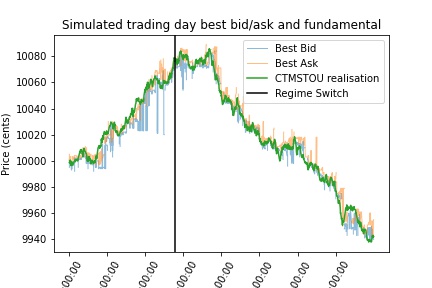} % Reduce the figure size so that it is slightly narrower than the column. Don't use precise values for figure width.This setup will avoid overfull boxes.
    \caption{Best bid \& Best ask for 1 simulated day ($\lambda = 2.90\ [event/day],\ \omega= 0.812\ [event/day]$.}.
    \label{fig:simulation_1_day}
    \end{figure}

    Once the parameters $\lambda$ and $\omega$ are calibrated we can use them for the CTMSTOU oracle. Running simulation episodes; figure \ref{fig:simulation_1_day} shows an example of a simulated trading day using the parameters obtained in section ~\ref{section:find_parameters}. The figure \ref{fig:simulation_1_day} shows an example of simulation output representing the market best bid and best ask throughout a day in which one regime switch happened.
    
    \section{Experimentation}
    We propose using the CTMSTOU process as a fundamental in the ABIDES-Markets simulator to illustrate the importance of \emph{regime-awareness} when deploying algorithmic-trading strategies. 
    \subsection{Detecting regime shifts }
    Many methods exists in the literature proposing to detect regime shifts. \cite{deadvances} surveys structural breaks detection methods such as CUSUM tests \cite{lee2003cusum}, explosiveness tests or sub/super martingale tests \cite{lopez2018advances}. Since our focus is demonstrating the importance of the \emph{regime-awareness}, we simply directly give our experimental trading agent access to the regime in real time. Concretely, at all time, the experimental agent in ABIDES-Markets can query the market oracle to know the current market regime: bearish or bullish trend.
    
    \subsection{Experiment Intuition}
    Our experiment demonstrates the importance of \emph{regime-awareness} in the classic problem of algorithmic optimal execution as detailed in \ref{sec:intro}. We propose to test four different algorithms. They all share the same quantity \emph{Scheduling} but have different \emph{Order Placement Models} (OPM). The first trade-off to consider when coming up with an OPM is the choice to place market orders (Instruction to buy or sell immediately at the current best available price) or limit orders (Buy or sell only at a specified price or better) \cite{tsai2007limit}. Intuitively, in the context of the execution problem without time limit, placing limit orders only is the best decision as it does not cross the spread and reduce the impact on the market. However, with a  time limit and particularly in the presence of market regimes, this conclusion does not hold: e.g placing a limit buy order in an upward trending market is risky. Indeed, the order may never be executed because the best bid price has already move-up, possibly even higher than the best ask at the previous time step: placing market orders would be a preferable choice here. On the other side, in a downward trending market, placing market buy orders is a less efficient choice. Since the market best ask price goes down, limit orders are more likely to be filled between two time periods resulting in a more efficient decision. We want to demonstrate that the \emph{regime-awareness} enables better decision making in the context of execution.

    \subsection{Proposed Protocol}
    We use ABIDES-Markets stylized background agent configuration modified using the CTMSTOU fundamental with the continuous-time Markov transition matrix parameters defined in figure \ref{fig:simulation_1_day}.  
    For each simulated trading day, an agent is tasked with a parent order consisting of buying a parent quantity $Q=20000$ shares before a time limit $T=23\ \textrm{hours}$. Their scheduling is based on the TWAP \ref{sec:intro} model; a time period $\tau=1\ \textrm{minute}$ is defined and each child order is of a size $q = \frac{q \times \tau}{T}$. The OPMs are different and are defined as such:
    \begin{itemize}
        \item \emph{full\_MO}: the agent places only market orders
        \item \emph{full\_LO}: the agent places only limit orders at the price level best bid
        \item \emph{regime\_aware\_0}: before placing an order, the agent queries the regime. If the market is in an increasing trend regime, the agent places a Market order, otherwise it places a limit order. The intuition is in an increasing market, an execution algorithm needs to secure the execution of the child order rather than risking further increase of the stock/coin price. In a decreasing trend market, an execution algorithm would place limit order; indeed price is decreasing so it may place a limit order and wait other market participants cross the spread towards the downward trend.
        \item \emph{regime\_aware\_1}: this OPM strategy tries to benefits even more from \emph{regime-awareness}. The scheduling is artificially slightly modified, the time period $\tau$ considered is increased by $k=10$ folds to $\tau=10 \ \textrm{minutes}$. In an upward market, the OPM places a market order of size $k \times q$, while in a downward market, the OPM layers $k$ limit orders of size $q$ between the levels $\{bestBid, \ ...,\ BestBid - k\}$. 
    \end{itemize}
    
We simulate 100 trading days (using 100 different simulation seed, everything else kept identical) for each execution strategy. For each strategy, we are interested in the average over the 100 simulated days of: 
\begin{itemize}
    \item Weighted average price $WAPr = \frac{\sum_{i=1}^{N}q_i*P_i}{\sum_{i=1}^{N}q_i}$ with $N$ the number of executed individual orders $\{o_i\}_{i=1..N}$, $q_i$ the executed quantity of order $o_i$ and $P_i$ its price. In the case of a buy execution parent order, we aim to minimize this quantity. 
    \item Percentage of completion $PctComp = \frac{\sum_{i=1}^{N}q_i}{Q}$ with $Q$ the size of the parent order. We aim to maximize this quantity, an efficient algorithm should ensure near full completion of the parent order execution.
\end{itemize}

    \subsection{Empirical results}
    Table \ref{tab:experiment_results} shows the average results for $WAPr$ and $PctComp$ over the 100 simulated trading days. We identify that \emph{regime\_aware\_0} ensures the best results in terms of pure average execution price, however, it does not ensure full execution of the parent order on average. \emph{Regime\_aware\_1} is the best strategy at minimizing the $WAPr$ while ensuring near completion of the parent order.  
    
\begin{figure}[t]
    \centering
    \includegraphics[width=0.7\columnwidth]{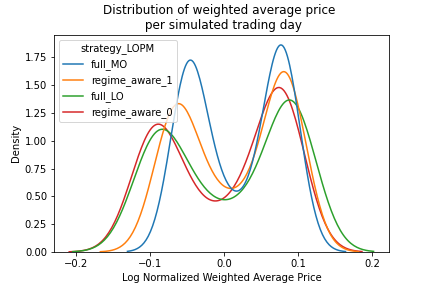} % Reduce the figure size so that it is slightly narrower than the column. Don't use precise values for figure width.This setup will avoid overfull boxes.
    \caption{Density of the weighted average price of execution during a simulated day.}.
    \label{fig:density_price}
    \end{figure}

\begin{figure}[ht!b!]
    \centering
    \includegraphics[width=0.7\columnwidth]{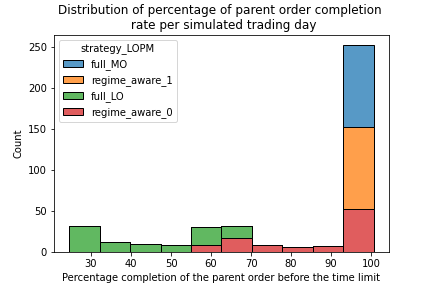} % Reduce the figure size so that it is slightly narrower than the column. Don't use precise values for figure width.This setup will avoid overfull boxes.
    \caption{Histogram of the percentage of completion of the parent order before the time limit.}.
    \label{fig:density_pct}
    \end{figure}

    Figure \ref{fig:density_price} reports the distribution of the $WAPr$ and \ref{fig:density_pct} reports the histogram of completion rate for the different algorithms. They reinforce previous conclusions. 
    
\begin{figure}[htb!]
    \centering
    \includegraphics[width=0.7\columnwidth]{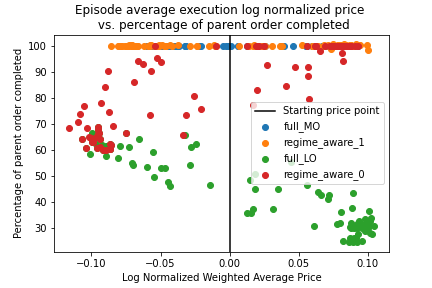} % Reduce the figure size so that it is slightly narrower than the column. Don't use precise values for figure width.This setup will avoid overfull boxes.
    \caption{Scatter plot of the completion of the parent order vs. average execution price.}.
    \label{fig:scatter}
    \end{figure}

    Figure \ref{fig:scatter} is particularly interesting as it confirms our original intuition about the trade-off market orders vs. limit orders in upward or downward regime markets. We observe a linear decreasing relationship between $WAPr$ and $PctComp$ for the \emph{full\_LO} algorithm. Indeed, a high $WAPr$ means that overall the market has been trending upward during the simulated day. The lower $PctComp$ here is due to limit orders not being filled between two time periods.
    
    Overall, this simple experiment has demonstrated that simple \emph{regime-aware} algorithms are able to outperform \emph{non-aware} algorithms. (Even though the use of regime-awareness by the policies in this work is relatively simple)

    \begin{table}[t]
    \centering
    %\resizebox{.95\columnwidth}{!}{
    \begin{tabular}{|
    >{\columncolor[HTML]{C0C0C0}}l |c|c|}
    \hline
    \multicolumn{1}{|c|}{\cellcolor[HTML]{9B9B9B}{\color[HTML]{000000} \textit{\textbf{OPM}}}} & \cellcolor[HTML]{9B9B9B}{\color[HTML]{000000} \textit{\textbf{\begin{tabular}[c]{@{}c@{}}Percentage \\ Completion\end{tabular}}}} & \cellcolor[HTML]{9B9B9B}{\color[HTML]{000000} \textit{\textbf{\begin{tabular}[c]{@{}c@{}}Average \\ Normalized\\  Price\end{tabular}}}} \\ \hline
    \textit{Regime Aware 0} & 86.51\% & 1.003787 \\ \hline
    \textit{Full LO} & 45.64\% & 1.012129 \\ \hline
    \textit{Regime Aware 1} & 99.96\% & 1.016589 \\ \hline
    \textit{Full MO} & 100.00\% & 1.019352 \\ \hline
    \end{tabular}
    \caption{Average percentage of completion vs average normalized execution price over the 100 simulated trading days.}
    \label{tab:experiment_results}
    \end{table}
    
\section{Conclusion}

Our contribution is in three-fold: (1) introducing the novel Continuous-Time Markov Switching Trending Ornstein-Uhlenbeck (CTMSTOU), (2) extension of the ABIDES-Markets simulator fundamental to study regime switching problems and (3) illustration of the importance of \emph{regime-awareness} for trading algorithm performance. 

We hope this work will encourage further research in the direction of regime detection in financial markets and the study of regime-awareness impact on policies to achieve financial tasks.

We think this work lays out the foundations for simulated environments enabling research on the topic.

\footnotesize
\textbf{Acknowledgements}

We would like to thank Nelson Vadori for his contributions and valuable advice.\\

\scriptsize
\textbf{Disclaimer}

This paper was prepared for informational purposes by the Artificial Intelligence Research group of JPMorgan Chase \& Co\. and its affiliates (``JP Morgan''), and is not a product of the Research Department of JP Morgan. JP Morgan makes no representation and warranty whatsoever and disclaims all liability, for the completeness, accuracy or reliability of the information contained herein. This document is not intended as investment research or investment advice, or a recommendation, offer or solicitation for the purchase or sale of any security, financial instrument, financial product or service, or to be used in any way for evaluating the merits of participating in any transaction, and shall not constitute a solicitation under any jurisdiction or to any person, if such solicitation under such jurisdiction or to such person would be unlawful.

\normalsize

% Use \bibliography{yourbibfile} instead or the References section will not appear in your paper
\bibliography{aaai22}

\appendix

\section{Appendix}

       \begin{algorithm}[!htb]
        \caption{Continuous-time transition matrix calibration}
        \label{alg:tr_mat}
        \textbf{Input}: $D$ set of real data, $C$ the number of trial in the search\\
        \textbf{Parameter}:$\kappa$ a method to count regime switches from OHLC (open, high, low, and closing prices for each period) data \cite{ohlc_investopedia}, $\Delta$ a method to measure distance between two distributions\\
        \textbf{Output}: a calibrated transition matrix $\Pi$\\
        \begin{algorithmic}[1] %[1] enables line numbers
        \STATE Let $D_{trans}$ the distribution of real data regime switches count an empty list.\\
        \COMMENT{Compute number of regime switches from real data}
        \FOR{$tradingDay \in D$}
        \STATE Compute $switchCount = \kappa(tradingDay)$
        \STATE Assign  $switchCount$ to $D_{trans}$
        \ENDFOR\\
        \COMMENT{Simulate Data}
        \STATE Let $S_{trans}$ the distribution of simulated data regime switches count an empty list.\\
        \STATE Let $c=0$,  $\delta_{best} = \infty$, $mu_{best} = \lambda_{best} = 0$
        \STATE Let $\delta_{best} = \infty$
        \WHILE{$c<C$}
        \STATE Draw $\omega \sim U(0,1)$, $\lambda \sim U(0,1)$
        \FOR{$i \in len(D)$}
        \STATE Simulate $tradingDay=\textrm{CTMSTOU}(\lambda,\omega)$
        \STATE Compute $switchCount = \kappa(tradingDay)$
        \STATE Assign  $switchCount$ to $S_{trans}$
        \ENDFOR
        \STATE Compute $\delta = \Delta(S_{trans}, D_{trans})$
        \IF {$\delta<\delta_{best}$}
        \STATE  $mu_{best} = \omega, \lambda_{best} = \lambda, \delta_{best}=\delta$
        \ENDIF
        \STATE $c\longleftarrow c+1$
        \ENDWHILE
        \STATE \textbf{return} $\Pi = \begin{pmatrix} \lambda_{best} & \omega_{best}\\ \omega_{best} & \lambda_{best} \end{pmatrix}$
        \end{algorithmic}
        \end{algorithm}

         \begin{algorithm}[!htb]
        \caption{Regime labeling algorithm for 1-min data}
        \label{alg:reg_lab}
        \textbf{Input}: $D$ set of real data for a day\\
        \textbf{Parameter}:$\tau_1$ time horizon of the short Moving Average, $\tau_2$ time horizon of the long Moving Average, $\alpha$ re-scaling coefficient\\
        \textbf{Output}: $N$ the number of regime switches\\
        \begin{algorithmic}[1] %[1] enables line numbers
        \STATE Let $x_{ref} = x_0$, $N=0$
        \FOR{$t \in D$}
        \STATE Compute $m_{\tau_1}=D_{MA_{\tau_1}}(t)$, $m_{\tau_2}= D_{MA_{\tau_2}}(t)$ and $s_{\tau_1} = D_{STD_{\tau_1}}(t)$
        \STATE Compute $x_t = \frac{m_{\tau_2} - m_{\tau_1}}{\alpha s_{\tau_1}}$
        \IF{$((x_{ref}>-1 )\land (x_{t}<-1))\lor ((x_{ref}<1)  \land (x_{t}>1))$}
        \STATE $N \longleftarrow N +1 $
        \ENDIF
        \STATE $x_{ref} \longleftarrow x_t$
        \ENDFOR
        \STATE \textbf{return} $N$
        \end{algorithmic}
        \end{algorithm}

\end{document}